# Open-Source Retrieval Augmented Generation Framework for Retrieving Accurate Medication Insights from Formularies for African Healthcare Workers

Joshua Owoyemi, PhD.
*ModAstera, Inc. Tokyo, Japan*

Shamsudeen Abubakar, PhD,
University of Louisville

Ayomide Owoyemi, MD, PhD
University of Illinois, Chicago

Taofeeq Oluwatosin Togunwa, MB;BS
University of Ibadan, Ibadan

Favour Chiemerie Madubuko, BSc
Ashesi University Ghana

Samuel Oyatoye, MS
Texas Tech University

Zeenat Oyetolu, BPharm
Advantage Health Africa

Kelvin Akyea, BSc
University of Cape Coast

Afeez Olamilekan Mohammed, Bsc
Obafemi Awolowo University, Ile-Ife

Abimbola Adebakin, BPharm, MBA
Advantage Health Africa

*Abstract*— Accessing accurate medication insights is vital for enhancing patient safety, minimizing errors, and supporting clinical decision-making. However, healthcare professionals in Africa often rely on manual and time-consuming processes to retrieve drug information, exacerbated by limited access to pharmacists due to brain drain and healthcare disparities. This paper presents "Drug Insights," an open-source Retrieval-Augmented Generation (RAG) chatbot designed to streamline medication lookup for healthcare workers in Africa. By leveraging a corpus of Nigerian pharmaceutical data and advanced AI technologies, including Pinecone databases and GPT models, the system delivers accurate, context-specific responses with minimal hallucination. The chatbot integrates prompt engineering and S-BERT evaluation to optimize retrieval and response generation. Preliminary tests, including pharmacist feedback, affirm the tool's potential to improve drug information access while highlighting areas for enhancement, such as UI/UX refinement and extended corpus integration.

## I. INTRODUCTION (*HEADING 1*)

Accessing accurate drug information is vital for safe medication use, improving patient outcomes, and minimizing errors [1]. Doctors and Pharmacists are crucial in ensuring patients receive accurate, relevant, and applicable medication and information. The ease of accessing relevant drug information is integral to quality care delivery and reducing the burden for Doctors and Pharmacists [2]. Moreover, timely access to accurate information helps identify potential contraindications and drug interactions, preventing adverse drug reactions and ensuring patient safety

Most doctors and pharmacists in Nigeria and Africa manually retrieve drug information using handbooks like Emdex or the British National Formulary, search engines like Google, and websites like WebMD to find the needed information [2,3]. While manual formularies provide comprehensive medication information, retrieving details from them is often laborious and time-consuming due to their extensive volume. Leveraging Large Language Models (LLM) to achieve this using Retrieval Augmented Generation (RAG) frameworks is possible.

RAGs couple sophisticated database retrieval mechanisms with advanced natural language processing (NLP) powered by LLMs to provide accurate and contextual responses with minimal hallucinations. These systems integrate structured and unstructured data, enabling seamless interactions and delivering results tailored to complex queries. They are gaining general use for enterprise [4,5] and academic applications [6,7] and increasing particular use in medical applications [8,9], where precision and reliability are critical. Their versatility has made them especially valuable for addressing knowledge gaps, automating workflows, and supporting decision-making across various industries.

This research was conducted to create an open-source framework to ease the retrieval of relevant medication information from any formulary used in Nigeria. By leveraging RAG technology, the framework aims to improve healthcare accessibility and foster better clinical outcomes through timely and accurate data access. Additionally, the system is designed to accommodate Nigeria's diverse formulary standards, ensuring compatibility and scalability for long-term use.

## II. METHODS

### A. Data Extraction

Textual data was extracted from a collection of drug-related PDF documents, including the EMDEX (Essential Medicines InDEX) complete Drug formulary [8], the most used drug and therapeutic information reference source by healthcare professionals in Nigeria. These documents are used as the foundational corpus of the drug insights RAG LLM. The text extraction process leveraged PyMuPDF[a1] for handling complex PDF layouts and document structures. This library was selected specifically for its ability to process multi-column formats, nested layouts, and documents containing mixed layout patterns while preserving the logical reading order. PyMuPDF's text extraction algorithm employs advanced heuristics to detect and adequately sequence content blocks, maintaining the semantic structure of the source material. The library's internal PDF parser identifies distinct layout regions through spatial analysis and content flow detection, enabling accurate text extraction even from documents with varying column widths, merged cells, and embedded figures. Using PyMuPDF's get_text() method with the "text" flag, we obtained raw text output while preserving paragraph boundaries and natural content flow. The library's built-in metadata extraction capabilities also facilitated the retrieval of document properties, including title and author, which provided valuable context for the subsequent analysis pipeline. This approach proved particularly effective for our corpus, where traditional linear text extraction methods often failed to correctly interpret the spatial relationships between content elements. Maintaining the Integrity of the Specifications

*B. Data Standardization and Pre-processing*

To standardize and structure the raw text data related to drugs, the data was first loaded and segmented into manageable chunks (batches). A structured data schema, defined by a primary Drug class, outlined essential details such as drug name, indications, contraindications, dosages, and side effects. The Azure-based GPT model (AzureChatOpenAI)[b] was then utilized to process each text chunk, ensuring the structured output was accurate and clear. A predefined prompt template guided the model to systematically refine and organize the drug information according to the schema. Finally, the structured data was appended to a new text file in the designated output directory. This approach effectively transformed unstructured drug data into organized, informative content suitable for further analysis and application.

*C. Vectorization and Storage*

The extracted structured text files in ".txt" format are processed and ingested into a Pinecone database[c], where each file is converted into tokens generated by the AzureOpenAIEmbeddings model[d]. These embeddings are high-dimensional vectors with a dimension size of 1536 chosen to optimally capture semantic meaning and contextual relationships within the texts. The Pinecone database uses cosine similarity with a threshold of 0.9 and similarity count of 3, during retrieval operations, ensuring efficient and accurate nearest-neighbor searches for vector similarity.

*D. Response Generation*

For response generation, the system leverages the capabilities of the GPT-4o model[e], which utilizes its advanced natural language understanding to generate highly coherent and contextually relevant outputs. The embeddings stored in Pinecone serve as the foundational knowledge base. At the same time, the GPT-4o model enhances the retrieval process by refining and contextualizing the retrieved information to produce detailed, human-like responses.

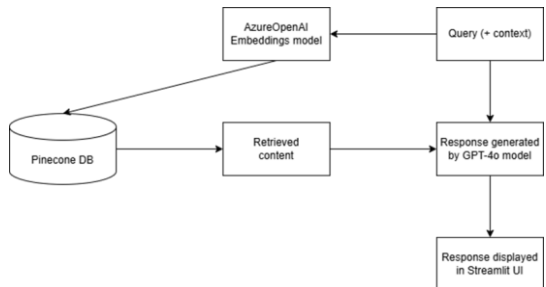

Fig. 1: Architecture of Drug Insights

## III. PROMPT ENGINEERING AND OUTPUT EVALUATION

*A. Prompt Engineering*

Prompt Artisan [11], a widely used custom GPT for co-creation of prompts using best practices, was used in the creation of the default QA system and context prompts suitable for a drug interaction assistant. These prompts are zero-shot style prompts. Using reflections on contemporary prompt engineering techniques such as those in [12], we created 9 prompts for evaluation using two classes generated response sentence limit and comparison/guardrails use as shown in Table 1. For example, prompt_0a is a prompt that has no sentence limit, compares 4 generated results before selecting one and guardrails which are explicit instructions to further reduce the chance of hallucination (no speculation, no unverified sources, clear disclaimer, context-limited responses).

*B. Output evaluation using S-BERT*

To evaluate the performance of 50 queries selected and reviewed by Doctors and Pharmacists under Drug effects, Dosage, Side effects and Special populations headings. Using a given field query, we compared similarity of Drug_insights response with the response from a pharmacist that used only the corpus accessible by Drug_insights. Similar human and drug insights responses were compared using Sentence-BERT (S-BERT), a variant of the BERT (Bidirectional Encoder Representations from Transformers). Designed for sentence-level tasks, S-BERT score measures the cosine similarity between sentence embeddings generated by S-BERT; with a higher score indicates a closer semantic relationship between two sentences. S-BERT scores are interpretable, scalable, and particularly effective in tasks requiring a fine-grained understanding of sentence-level semantics [13]. As a check, a few questions whose answers were not present in the corpus provided were included in the query set.

The output of drug insights with each of these prompts to each query in the test question set were then compared with the response by the pharmacists and the results are shown in Fig 3.

---

a: PyMuPDF is a high-performance Python library for data extraction, analysis, conversion & manipulation of PDF (and other) documents. (https://pymupdf.readthedocs.io/en/latest/)
b: AzureChatOpenAI is a Microsoft Azure service that provides powerful language models from OpenAI
c: Pinecone is a fully managed vector database that makes it easy to add semantic Search. (www.pinecone.io)
d: AzureOpenAIEmbeddings model: https://github.com/langchain-ai/langchain/blob/master/libs/community/langchain_community/embeddings/azure_openai.py
e: GPT-4o model is a multimodal large language model that supports real-time conversations, Q&A, text generation and more

| Prompt class | Designation | Description |
|---|---|---|
| Class 1 (generated output length) | prompt_*a | No limit |
| | prompt_*b | Limit to 2 sentences |
| | prompt_*c | Limit to 3 sentence |
| Class 2 (Compare and guardrails) | prompt_0* | Compare 4 generated results + guardrails |
| | prompt_1* | guardrails |
| | prompt_2* | Compare |

Table 1: Classes of tested prompts

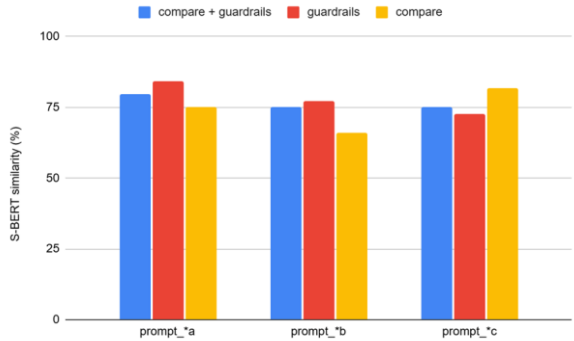

Fig. 2: Average S-BERT similarity score (in %) between Drug insights responses with different prompts and human pharmacists.

From Fig 2, we can see that prompt_0a - a template prompt with guardrails, no comparison of multiple generations, and no limit in generated response had the best similarity score (84.09%) and prompt_1b had the worst (65.91%). There is no direct linear mapping between changes and scores, which is typical in prompt engineering.

*C. Output evaluation using Human evaluators*

The app was evaluated by pharmacists affiliated with Advantage Health Africa, a healthcare solution company providing technology products and services to promote affordable and quality healthcare across Africa [14]. They were provided with queries to and responses from the app and evaluated along several usability, accuracy questions on a 1-5 agreement scale where 1 strongly disagrees and 5 is mostly agree.

| Evaluation question | Average score (/5) |
|---|---|
| The output relevant to the question | 3.88 |
| The output is accurate | 3.76 |
| The output is usefully constructed | 3.88 |
| The added source documents are relevant | 3.76 |

Table 2: Pharmacist feedback scores on relevancy, accuracy and display of responses by Drug insights

## IV. PREPARE HOW IT WORKS, INCLUDING ARCHITECTURE AND PERFORMANCE METRICS

The system is a RAG chatbot designed to mitigate hallucinations during response generation. It leverages a corpus of leading pharmaceutical data from Nigeria for precise, context-aware outputs. The architecture integrates various technologies: Pinecone is used as the database for efficient vector storage and retrieval; Azure ChatGPT is employed as the language model (LLM) engine for generating accurate responses; and LangChain API connects these components, ensuring smooth interaction between the LLM and the database. Streamlit provides a user-friendly interface, making the tool accessible and interactive. The system also supports easy configuration for multiple prompts, enhancing adaptability for different use cases.

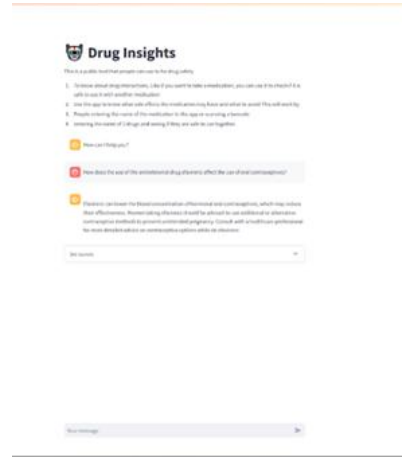

Fig. 3: Image of Drug Insights with a sample query and response

## V. FUTURE ENHANCEMENTS OR EXPANSIONS

Future areas for potential improvement include:
- System testing:
  - Explicit systematic evaluation of the recall and precision of database retrievals.
  - Current tests on Drug Insights were done with one-shot questions. Tests that evaluate performance over a chat session would be useful.
  - Consider other human-LLM similarity metrics besides S-BERT.
- UI/UX improvements:
  - Include ways to capture real-time user feedback data, such as like/dislike button - as a way to improve performance
  - Refine display of references in the UI such as including page number.

## VI. CONCLUSION

Drug Insights demonstrates the transformative potential of Retrieval-Augmented Generation (RAG) in healthcare by providing accurate and accessible drug information tailored to the needs of Nigerian healthcare professionals. Leveraging advanced natural language processing, robust database retrieval, and prompt engineering, the system delivers highly accurate contextually relevant insights, as validated by S-BERT metrics and pharmacist feedback. While the tool effectively addresses current gaps in drug information access, future enhancements such as expanded testing, refined user interfaces, and real-time feedback integration will further optimize its performance. This open-source innovation lays a solid foundation for scalable, data-driven solutions in healthcare.

*The Github Repo is open source and available at https://github.com/OpenSourceCollective/drug_insights*

ACKNOWLEDGMENT

We want to acknowledge Advantage Health Africa and other members of Axum AI for their support in carrying out this research